\documentclass{PoS}

\usepackage{textcomp}
\usepackage{amsfonts} 
\usepackage{amssymb} 
\usepackage{amsmath} 
\usepackage{slashed}
\usepackage{graphicx}
\usepackage{pifont}
\usepackage{multirow,lscape}
\usepackage{array}
\usepackage{longtable}
\usepackage[FIGBOTCAP,TABBOTCAP,tight]{subfigure}
\usepackage{verbatim}
\usepackage{bm}
\usepackage{comment}
\usepackage{xcolor}
\usepackage{url} 
\usepackage{epstopdf}
\usepackage{grffile}
\usepackage{multicol}
\usepackage{simplewick} 
\usepackage{braket}
\usepackage{cancel}

\newcommand{\wtd}{\widetilde}

\graphicspath{{./figs/}}

\title{Towards a determination of the nucleon EDM from the quark chromo-EDM operator with the gradient flow}

\ShortTitle{Quark-chromo EDM with gradient flow}

\author{\speaker{Jangho Kim}, Jack Dragos, Andrea Shindler\\
      Facility for Rare Isotope Beams, Physics Department, Michigan State University, East Lansing, Michigan, USA \\
        E-mail: \email{kimjangho@nscl.msu.edu}
}

\author{
        Thomas Luu\\
        Institute for Advanced Simulation (IAS-4) FZJ, Germany
}
\author{
        Jordy de Vries\\
        Amherst Center for Fundamental Interactions, Department of Physics, University of Massachusetts Amherst, Amherst, MA, USA\\
        RIKEN BNL Research Center, Brookhaven National Laboratory, Upton, New York, USA
}
\abstract{
In this proceedings, we lay the foundation for computing the contribution of quark chromo-electric dipole moment (qCEDM) operator to the nucleon electric dipole moment. 
By applying the gradient flow technique, we can parameterize the renormalization
and operator mixing issues associated with the qCEDM operator on the lattice.
As the nucleon mixing angle $\alpha_N$ is a key component for determining the neutron and proton
electric dipole moments induced by the qCEDM operator, we present the formalism and preliminary results for $\alpha_N$
with respect to the gradient flow time $t_f$.
The results are computed on $N_f=2+1$ Wilson-clover lattices provided by PACS-CS~\cite{Ishikawa:2007nn}.
The 3 ensembles have lattice spacing values of $a=\lbrace 0.1095,\,0.0936,\,0.0684\rbrace $~fm,
whilst keeping a similar $m_{\pi}\approx701$~MeV, and a fixed box size of $L\approx1.9$~fm.
}

\FullConference{The 36th Annual International Symposium on Lattice Field Theory - LATTICE2018\\
		22-28 July, 2018\\
		Michigan State University, East Lansing, Michigan, USA.}

\begin{document}
\section{Introduction}
A nonzero neutron electric dipole moment (nEDM) would indicate that CP symmetry is violated. The current bounds on the nEDM are a few orders of magnitude
above what can be induced via the weak sector and any non-zero signal can be attributed to new sources of CP violation. Apart from the theta term,  the Standard Model does not provide such CP-violating terms and the nEDM can provide evidence for physics beyond the Standard Model (BSM).
In lattice QCD, we explicitly compute the neutron and proton EDM induced by individual terms in the QCD action.
In these proceedings, we compute a crucial component of the nEDM, the nucleon mixing angle $\alpha_N$,
which induced by the quark chromo-electric dipole moment (qCEDM), an effective CP-violating BSM operator.
Renormalization of the qCEDM poses a challenge, as it mixes with operators of the same and lower dimension. By employing the gradient flow~\cite{Luscher:2013cpa}
to the qCEDM operator for varying flow time parameter $t_{f}$, we hope in the future to disentangle the mixing.
\section{Lattice Parameters}
\begin{table}
\caption{
\label{tab:lattice}
Summary of important lattice parameters for our ensembles, which share a common ${m_{\pi}\approx701}$~MeV
 and $L\approx1.9$~fm. The computing of $a$ and $m_{\pi}$ was performed in \cite{Ishikawa:2007nn}.
}
\begin{center}
\begin{tabular}{c | c | c | c | c | c | c }
\hline
\hline
$L_s^{3} \times L_t$& $a[fm]$  &$\beta$  & $\kappa_l$ & $\kappa_s$ & $c_{SW}$ & Nconfs \\
\hline
 $16^3 \times 32$  &0.1095(25) &1.83 &  0.13825 & 0.13710 & 1.761 & 800 \\
 $20^3 \times 40$  &0.0936(33) &1.9  &  0.13700 & 0.13640 & 1.715 & 790 \\
$28^3 \times 56$  &0.0684(41) &2.05 &   0.13560 & 0.13510 & 1.628 & 650 \\
\hline
\hline
\end{tabular}
\end{center}
\end{table}
To explore the discretization effects of $\alpha_{N}$, we use three PACS-CS lattice ensembles of varying lattice
 spacings, obtained through the ILDG~\cite{Beckett:2009cb}.
Having $N_{f}=2+1$, they were computed with a Iwasaki gauge action, and were used with a non-perturbative
O(a)-improved Wilson fermion action. All 3 ensembles exhibit similar box lengths of $L\approx 1.9$~fm and
similar pion masses of $m_\pi \approx 701 $~MeV. The computation within \cite{Ishikawa:2007nn} determined the
lattice spacing and pion masses, which has been summarized in Table.~\ref{tab:lattice}
\section{The qCEDM and the Two-Point Correlator}
\begin{figure}[htpb]
\subfigure[
\label{fig:3pt_conn}
Connected diagrams of three-point functions.
]{
\includegraphics[width=0.4\textwidth]{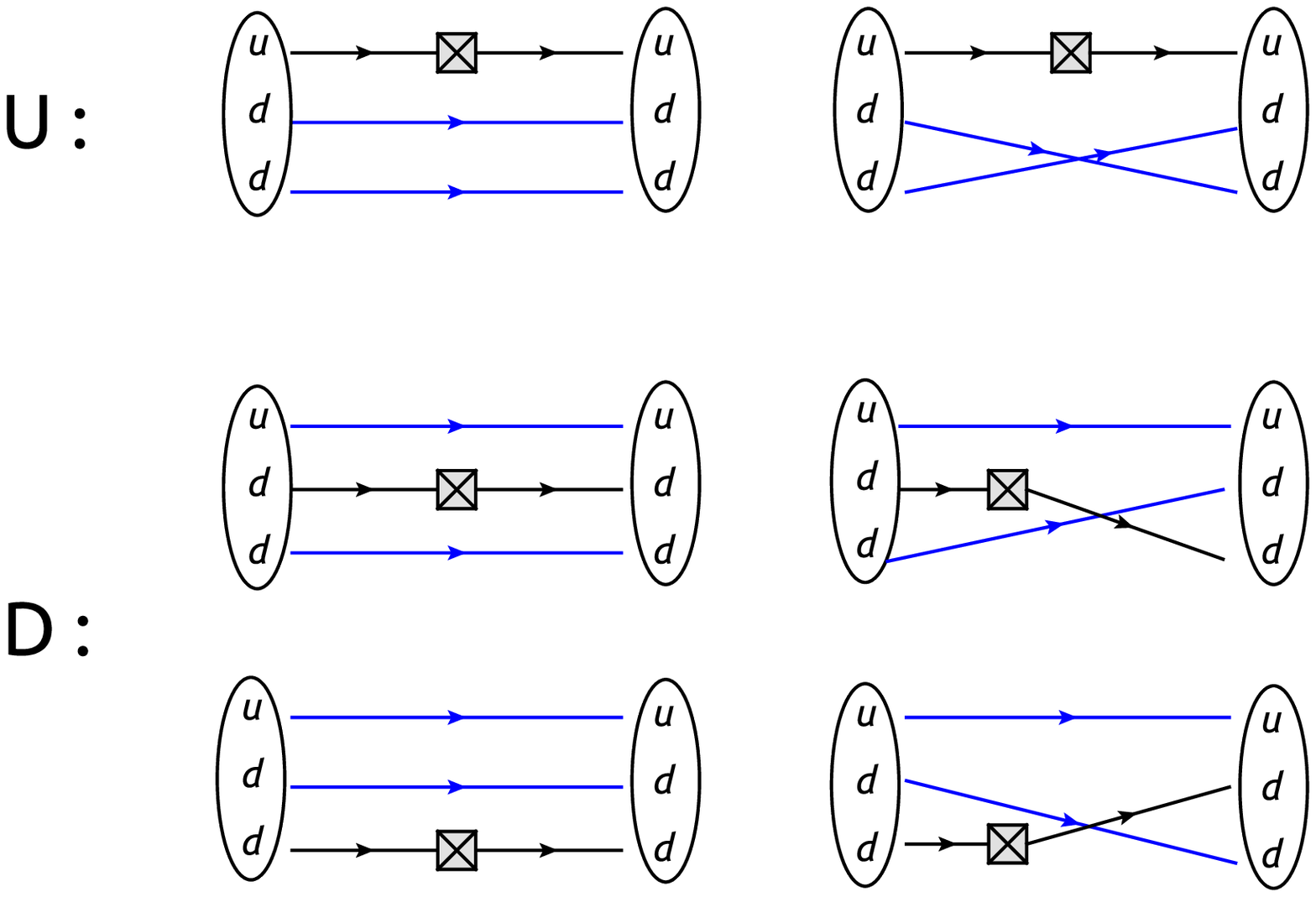}
}
\qquad \qquad
\subfigure[
\label{fig:3pt_disc}
Disconnected diagrams of three-point functions.
]{
\includegraphics[width=0.4\textwidth]{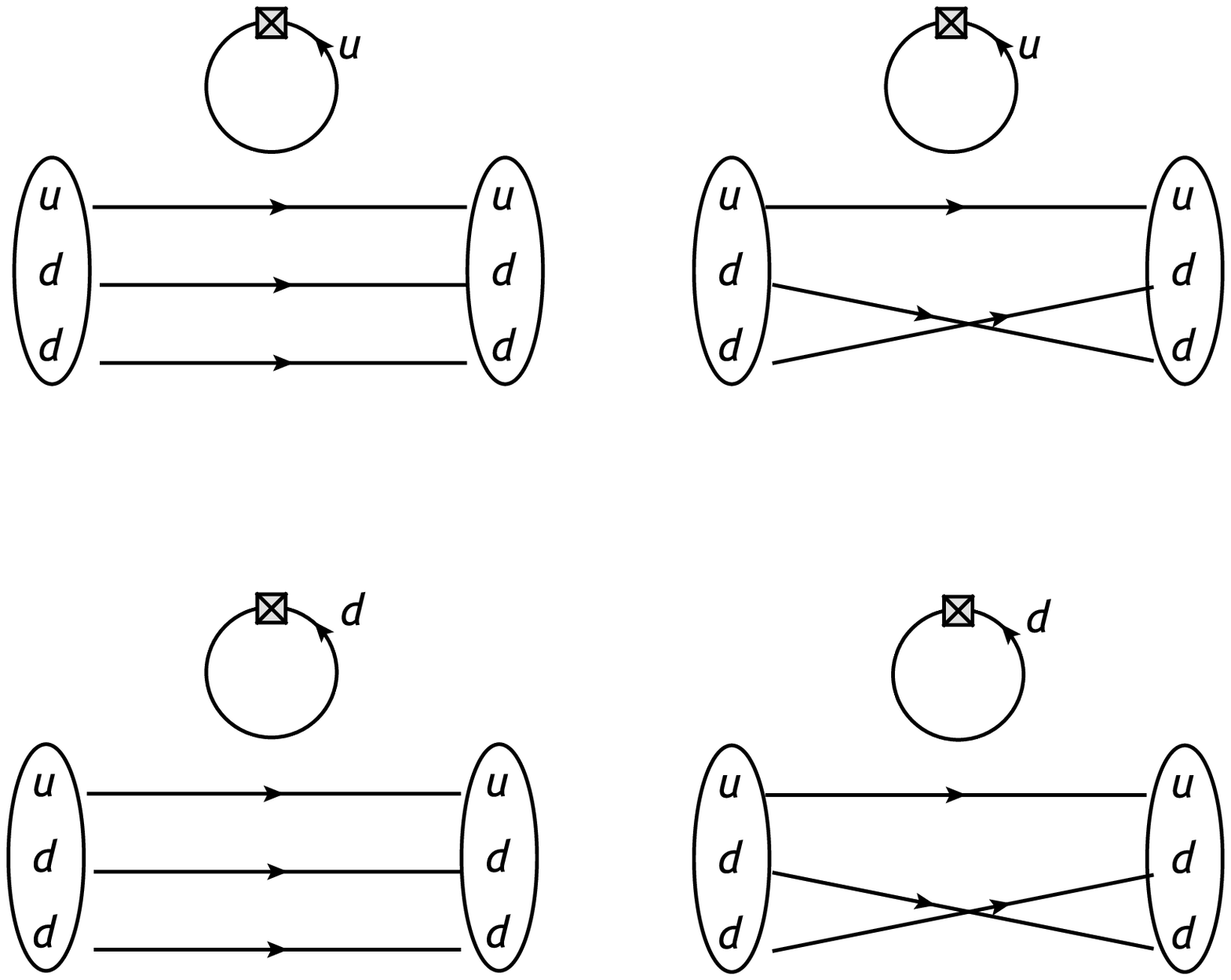}
}
\caption{
All 10 diagrams that contribute to the three-point correlation function.
The square box with cross symbol represents the qCEDM operator and the symbols $U:$ and $D:$ distinguish between
the terms were the qCEDM operator is inserted between $u$ or $d$ quarks.
Blue lines are sequential quark lines.
We do not compute the disconnected contribution in Fig.~\ref{fig:3pt_disc} for this proceeding.
}
\end{figure}

The qCEDM operator is defined as the particular bilinear combination:
\begin{align}\label{eq:qCEDM}
O(\vec{w},\tau) &= \sum_{f} [\bar{\psi}^{f}(\vec{w},\tau)]^{k}_{\kappa}
\Gamma^{kl}_{\kappa \lambda}(\vec{w},\tau) [\psi^{f}(\vec{w},\tau)]^{l}_{\lambda} \quad , \quad
\Gamma^{kl}_{\kappa\lambda}(\vec{w},\tau)= \frac{1}{2}\sigma_{\mu\nu} \gamma_5 [G_{\mu\nu}(\vec{w},\tau)]^{kl}_{\kappa\lambda}
\end{align}
where we have summed over all quark flavors $f$, dirac indices $\mu\&\,\nu$, color indices $k\&\,l$ and
spin indices $\kappa\&\,\lambda$. Unlike other common quark bilinears (e.g. currents, twist-two bilinears, etc...),
the qCEDM is a quark bilinear which also depends on the gluon field strength tensor $G_{\mu\nu}$.
We use the standard interpolating operators that have the quantum numbers of the neutron
\begin{align}
\bar{N}_{\gamma'}(\vec{0},0)
&=\epsilon^{a'b'c'} \bar{d}^{a'}_{\alpha'}(\vec{0},0) \wtd{C}_{\alpha' \beta'} \bar{u}^{b'}_{\beta'}(\vec{0},0) \bar{d}^{c'}_{\gamma'}(\vec{0},0), \nonumber\\
N_{\gamma}(\vec{x},t)
&=\epsilon^{abc} u^{a}_{\alpha}(\vec{x},t) \wtd{C}_{\alpha \beta} d^{b}_{\beta}(\vec{x},t) d^{c}_{\gamma}(\vec{x},t),
\end{align}
where $a,b,c, a', b', c'$ are color indices and $\alpha,\beta,\gamma,\alpha',\beta',\gamma'$ are spin indices.
We have define the combination $\wtd{C} = C \gamma_5$, where $C$ is charge conjugation matrix.
Using these interpolating operators, the two-point correlation function can be written in terms of the
quark fields
\begin{align}
G_{2}(P,t)=&
\sum_{\vec{x}}P_{\gamma'\gamma}
\bigg \langle
\left[
    \epsilon^{abc} u^{a}_{\alpha}(\vec{x},t)
    \wtd{C}_{\alpha \beta} d^{b}_{\beta}(\vec{x},t) d^{c}_{\gamma}(\vec{x},t)
    \right]
  \left[
      \epsilon^{a'b'c'} \bar{d}^{a'}_{\alpha'}(\vec{0},0) \wtd{C}_{\alpha' \beta'} \bar{u}^{b'}_{\beta'}(\vec{0},0) \bar{d}^{c'}_{\gamma'}(\vec{0},0)
      \right]
  \bigg \rangle \,.
\end{align}
For the construction of $\alpha_{N}$, we project out the positive parity nucleon states by setting \\
 ${P=T^{+}=\dfrac{I+\gamma_4}{2}}$.
We define our three-point correlation function as the two-point correlator with an insertion of the qCEDM operator which
is summed over space-time
\begin{align}
G_{3}(P,t) &=
\sum_{\vec{x},\vec{w},\tau}P_{\gamma' \gamma}\bigg \langle
\left[
  \epsilon^{abc} u^{a}_{\alpha}(\vec{x},t)
  \wtd{C}_{\alpha \beta} d^{b}_{\beta}(\vec{x},t) d^{c}_{\gamma}(\vec{x},t)
\right]
O(\vec{w},\tau)
\left[
  \epsilon^{a'b'c'} \bar{d}^{a'}_{\alpha'}(\vec{0},0) \wtd{C}_{\alpha' \beta'} \bar{u}^{b'}_{\beta'}(\vec{0},0) \bar{d}^{c'}_{\gamma'}(\vec{0},0)
\right]
\bigg \rangle\,.
\end{align}
By selecting $P=T^{+}\gamma_5$ for this correlation function (and $P=T^{+}$ for the two-point correlator), we
ensure the non-zero contribution to the states present in the spectral decomposition are consistent.
By virtue, this ensures that the ground state of the two- and three- point correlation functions have the same energy.
When performing the Wick contractions for the three-point correlation function, 10 independent terms are present.
The results shown in this paper only include the 6 ``connected diagrams'' shown in Fig.~\ref{fig:3pt_conn}, for
which the qCEDM is inserted on a propagator associated with the nucleon. We exclude the contributions coming from the
4 ``disconnected diagrams'' shown in Fig.~\ref{fig:3pt_disc}, for which the qCEDM operator is inserted on a quark propagator loop
that is disconnected from the propagators associated with the nucleon.
By performing a spectral decomposition of the two- (n=2) and three- (n=3) point correlation functions, the
``effective mass'' functions can be used to determine the mass of the nucleon
\begin{align}\label{eq:spec_decomp}
G_{n=2,3}(t) = Z_{n=2,3}e^{-M_{\text{eff} } t} \left[1 + O(e^{-\Delta Mt})\right] \quad , \quad
aM_{\text{eff}(n=2,3)}(t)=\log\left[\frac{G_{n=2,3}(t)}{G_{n=2,3}(t+1)}\right],
\end{align}
where $\Delta M$ is the difference between the first excited state and the ground state of the nucleon.
The region in which a plateau has occurred in the effective mass at large $t$ allows us to extract the
ground-state mass $M_{\text{eff}}$.
To average over the statistical noise, we fit the effective mass function over a region of $t$ beginning at where
the data has plateaued, and ending where the noise is starting to dominate over the signal.
\begin{figure}[htpb]
  \centering
\subfigure[
Effective mass function computed with the 2pt (green) and 3pt (red) functions.
]{
\label{fig:Eff_28X56}
\includegraphics[trim={5mm 0cm 2mm 0cm},width=0.45\textwidth]{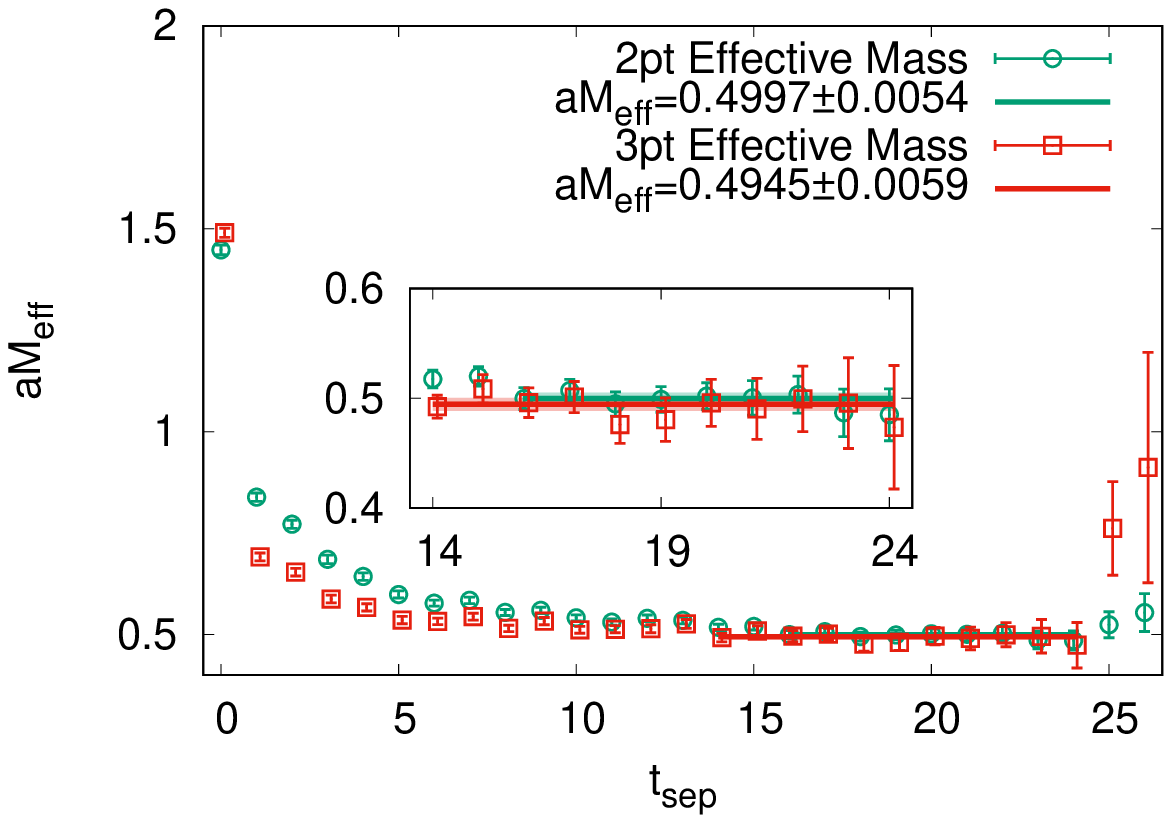}
}
\qquad
\subfigure[
Up quark (violet) and down quark (cyan) contributions to the nucleon mixing angle $\alpha_N$.
]{
\label{fig:alpha}
\includegraphics[trim={5mm 0cm 2mm 0cm},width=0.45\textwidth]{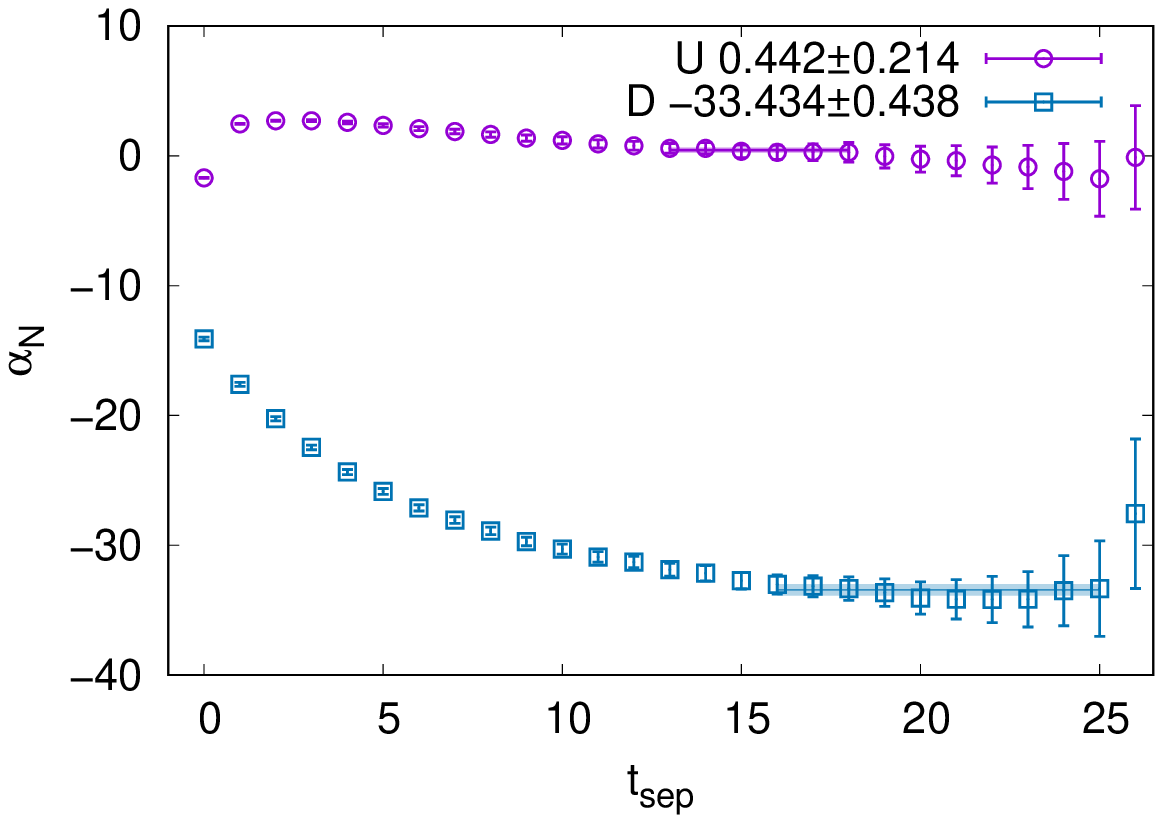}
}
\caption{
Effective mass function $M_{\text{eff}}(t_\text{sep})$ and nucleon mixing angle $\alpha_N$ computed on the $a=0.0684$ fm,
$28^3\times56$ lattice ensemble.
The bands in the left figure correspond to a constant fit in a region where the data points have plateaued to the
ground state.
}
\end{figure}
In Fig.~\ref{fig:Eff_28X56}, we compare the effective mass function computed using the two-point function (green),
with the effective mass computed using the three-point function (red) computed on the $28^{3}\times 56$ lattice.
In the region where both correlation functions have plateaued to the ground state, the data points
and their respective fit bands agree within the statistical error.
We observe that the effective mass function of the three-point correlator reaches
a plateau at a shorter source-sink separation $t_\text{sep}$ compared to the two-point function.
This indicates that the excited-state contamination effects (e.g. the $O(e^{-\Delta Mt})$ term in \ref{eq:spec_decomp})
are being suppressed simply by summing over the insertion of a qCEDM operator.
The nucleon mixing angle $\alpha_N$ provides a measure of how to ``rotate'' the nucleon spinnors
of the standard CP conserving theory, to a theory that includes a CP-violating term (in this case, the qCEDM).
To compute the nucleon mixing angle $\alpha_N$, we construct the ratio between the three- and two- point correlation functions
${\alpha_N (t_{\text{sep}}) \equiv -G_{3}(t_{\text{sep}})/G_{2}(t_{\text{sep}})}$,
which plateaus to $\alpha_N$ in the large $t_{\text{sep}}$ limit. Full derivations can be found in
Refs.~\cite{Abramczyk:2017oxr,Alexandrou:2015spa,Shintani:2015vsx,Shindler:2015aqa}.
In Fig.~\ref{fig:alpha}, we show the individual contributions
coming from the $U$ (violet) and $D$ (cyan) diagrams (defined in Fig.~\ref{fig:3pt_conn}) to $\alpha_N (t_{\text{sep}})$.
The figure clearly demonstrates that sum of the $D$ diagrams (cyan) are the major contributor to $\alpha_N$.
\section{Gradient Flow applied to the qCEDM}
\begin{figure}[htpb]
\center
\includegraphics[trim={5mm 0cm 2mm 0cm},width=0.7\textwidth]{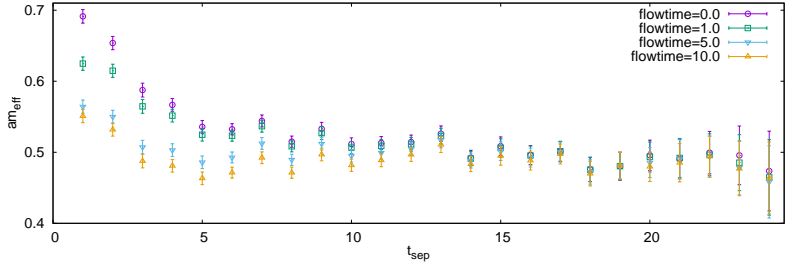}
\caption
{
\label{fig:flow_effective}
Effective mass function calculated from the flowed three-point correlators, computed on the ${a=0.0684}$~fm,
$28^3 \times56$ lattice ensemble. Different flow times of $t_{f}/a^{2}=\{0,1,5,10\}$ are plotted in purple, green, blue, orange
respectively.
}
\end{figure}

In this section, we repeat the extraction process for $M_{eff}$ and $\alpha_N$ as in the previous section,
but applying the gradient flow to the qCEDM operator within the two- and three- point correlation functions.
When following the gradient flow formalism from \cite{Luscher:2013cpa}, the quantity being flowed
(both for the gauge and/or fermion fields) is parameterized by the ``flow time" dimension $t_{f}$.
Since $t_f$ has a dimensionality of 2, the dimensionless combination of $t_f/a^2$ is the analogous parameter
on the lattice. As the graident flow has the effect of smearing the quark and gauge fields over space-time, the
``flow radius" $\sqrt{8t_{f}}/a$ provides a measure of the root-mean-square smearing radius for the
gradient flowed object.
The two terms that the gradient flow needs to be applied on in the qCEDM are the gluonic field strength tensor and the
quark fields. For the gluonic field strength tensor $G_{\mu\nu}(\vec{w},\tau)$, we apply the standard
gradient flow prescription for the gauge fields (see \cite{Luscher:2013cpa}), which are used to construct
the field strength tensor $G_{\mu\nu}(\vec{w},\tau,t_{f})$ at arbitrary $t_{f}$.
For the quark fields, it is more beneficial in lattice QCD to formalize the flow equations in terms of quark propagators
\begin{align}
S(t_f,\vec{y},t;\vec{x},t_0)=&\sum_{\vec{v},\tau}K(t_f,\vec{y},t;0,\vec{v},\tau)S(\vec{v},\tau,\vec{x},t_{0}) , \\
S(\vec{y},t;s_f,\vec{x},t_0)=&\sum_{\vec{w},\tau}S(\vec{y},t,\vec{w},\tau) K(s_f,\vec{x},t_0;0,\vec{w},\tau)^{\dagger}\, ,
\end{align}
where $K$ and $K^{\dagger}$ are flow operators which satisfies the quark flow equations
In total, the qCEDM operator that has been flowed has the form:
\begin{align}
  O(t_{f},\vec{w},\tau) &= \sum_{\substack{f,\vec{w}_{1},\vec{w}_{2}, \\ \tau_{1},\tau_{2}}}
  [\bar{\psi}^{f}(\vec{w}_{1},\tau_{1})]^{k}_{\kappa} K(t_{f},\vec{w},\tau;0,\vec{w}_{1},\tau_{1})^{\dagger}
  \Gamma^{kl}_{\kappa \lambda}(\vec{w},\tau)
  K(t_f,\vec{w},\tau;0,\vec{w}_{2},\tau_{2})  [\psi^{f}(\vec{w}_{2},\tau_{2})]^{l}_{\lambda}\,.
\end{align}
Since the gradient flow method is only applied to the qCEDM operator, the following three-point
function is computed
\begin{align}
G_3(P,t_f,t) =&\sum_{\vec{w},\tau,\vec{x}}
\ \ \sum_{\vec{w}_{1},\tau_{1}}
\sum_{\vec{w}_{2},\tau_{2}}
Tr\bigg[ PS_{\text{seq}}(\vec{x},t,\vec{0},0) \nonumber \\
&S(\vec{x},t,\vec{w}_{1},\tau_{1})
K(t_f;\vec{w},\tau,\vec{w}_{1},\tau_{1})^{\dagger}
\Gamma(t_f,\vec{w},\tau)
K(t_f,\vec{w},\tau;\vec{w}_{2},\tau_{2})
S(\vec{w}_{2},\tau_{2},\vec{0},0) \bigg]\,,
\end{align}
where $O(t_f,w)$ and $\Gamma(t_f,\vec{w},\tau)$ are defined in eq.~\ref{eq:qCEDM}, and
$S_{\text{seq}}(\vec{x},t,\vec{0},0)$ is collection of contracted propagators that are spectators to the current inserted
quark propagator line (blue lines in Fig.~\ref{fig:3pt_conn}). As for the unflowed three-point correlator, the projector
$P = T_{+}\gamma_{5}$ is selected.
To begin with, Fig.~\ref{fig:flow_effective} shows the effective mass function constructed from three-point
correlators, computed for different flow times of the qCEDM operator.
By increasing the flow time, we see the ground-state plateau region occurs at smaller source-sink separations $t_\text{sep}$.
This implies that the effect of applying larger flow times of gradient flow,
increases the relative size of the ground-state matrix elements to the excited-state matrix elements.
\begin{figure}[htpb]
\subfigure[
flow time $t_f/a^2=0.1$
]{
\label{fig:flowtime_0.1}
\includegraphics[trim={5mm 0cm 2mm 0cm},width=0.45\textwidth]{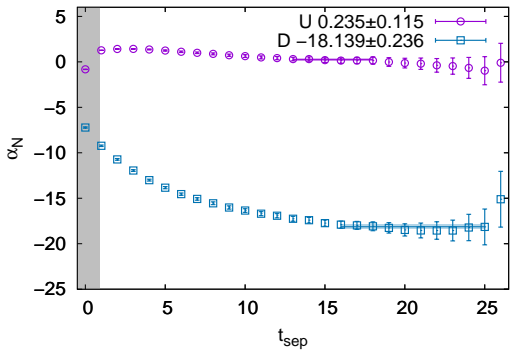}
}
\qquad
\subfigure[
flow time $t_f/a^2=10.0$
]{
\label{fig:flowtime_10.0}
\includegraphics[trim={5mm 0cm 2mm 0cm},width=0.45\textwidth]{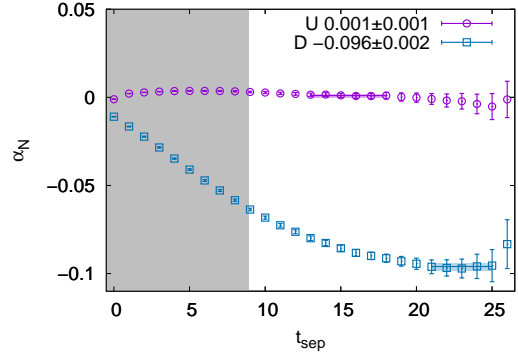}
}
\caption{
\label{fig:flowtime_alpha}
Flow time dependence of the flowed nucleon mixing angle $\alpha_N$, computed on the $a=0.0684$ fm, $28^3 \times56$ lattice ensemble. The Gray regions show where the source-sink separation $t_{\text{sep}}$ is smaller than the flow radius $\sqrt{8t_{f}}$.
}
\end{figure}
In Fig.~\ref{fig:flowtime_alpha}, we present the $U$ (violet) and $D$ (cyan) contributions to $\alpha_N$,
both computed at flow times $t_f/a^2=0.1$ (left) and $t_f/a^2=10.0$ (right).
The gray regions, in which the $t_{\text{sep}}$ is less than the flow radius $\sqrt{8t_{f}}$,
 should be excluded as the qCEDM operator extends to cover the whole systems time extent $t_{\text{sep}}$.
We firstly observe that the flow time dependence is greater effecting the $D$-diagrams contribution compared to the
 $U$-diagrams contribution.
Then, when flow time is increased, we observe that the $D$-diagram contribution to $\alpha_{N}$ has a plateau region beginning at larger values of $t_{\text{sep}}$.
\begin{figure}[htpb]
\subfigure[
$U$-diagrams contribution to $\alpha_N$
]{
\label{fig:alpha_U}
\includegraphics[trim={5mm 0cm 2mm 0cm},width=0.45\textwidth]{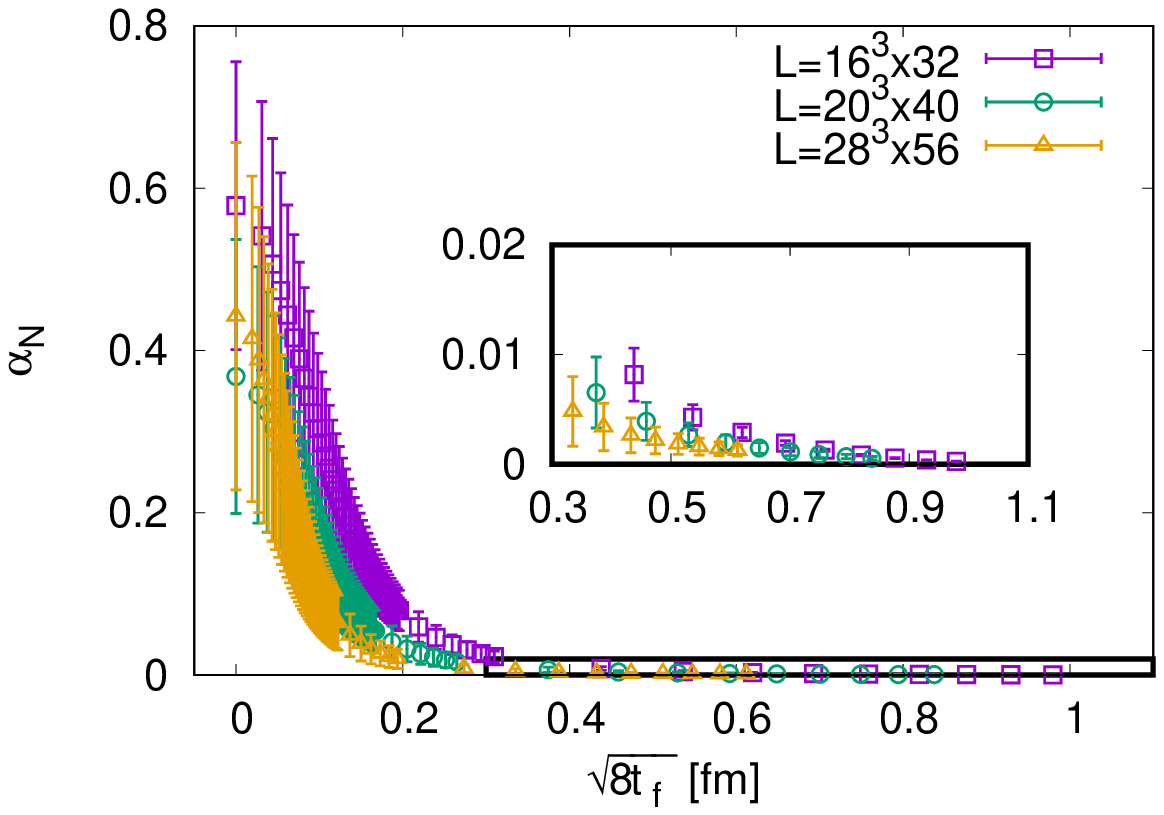}
}
\qquad
\subfigure[
$D$-diagrams contribution to $\alpha_N$
]{
\label{fig:alpha_D}
\includegraphics[trim={5mm 0cm 2mm 0cm},width=0.45\textwidth]{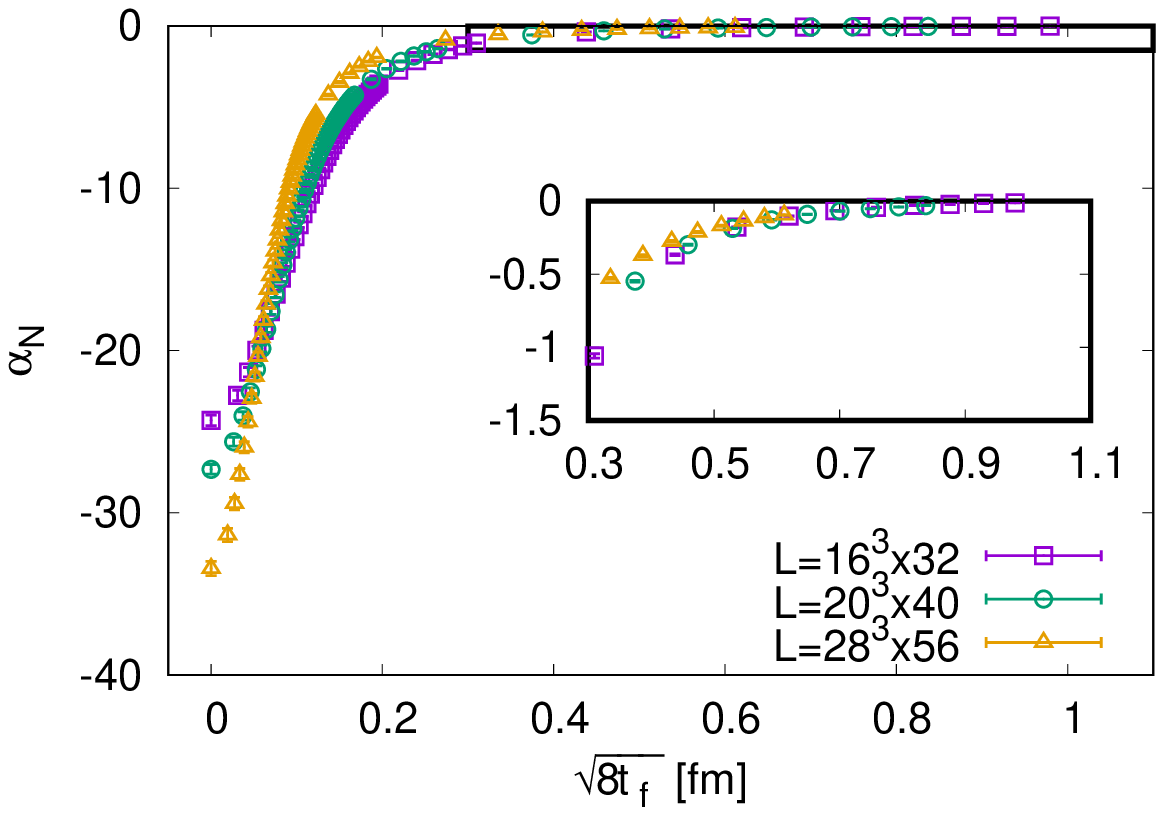}
}
\caption{
$\alpha_N$ plotted with respect with physical unit of flow time $\sqrt{8t_{f}}$ (in fm),
for the $L=16,\,20,\,28$ ensembles in violet, green and yellow.
\label{fig:flowtime_alpha_phy}
}
\end{figure}
In Fig.~\ref{fig:flowtime_alpha_phy},
we present the flow time dependence of the $U$ (left) and $D$ (right) contributions to $\alpha_N$,
with respect to the flow time radius $\sqrt{8t_f}$ converted to fm. The
$L=16^{3}\times 32$, $20^{3}\times 40$, $28^{3}\times 56$ ensembles are plotted in purple, green and orange.
We observe statistical agreement between all 3 lattice ensembles in the region $\sqrt{8t_f} > 0.6fm$.
\section{Conclusion}
In this proceedings, we presented our analysis of the effective mass and the
nucleon mixing angle computed using a three-point correlation function of which
the current operator is the qCEDM. 
By applying the gradient flow to the qCEDM, we explored how the effective mass
and nucleon mixing angle depends on the flow time parameter. 
As a result, we obtained an understanding of how the flow time radius
$\sqrt{8t_f}$ and the nucleon source-sink separation $t_{\text{sep}}$ can
affect each other. 
In addition, determining the flow time dependence of $\alpha_{N}$ will be
crucial for disentangling the operator mixing and performing the overall
renormalization of the qCEDM, within the gradient flow framework. 
Equipped with the flow time dependence of $\alpha_{N}$,
work has begun on computing the four-point functions, which will be used in
determining the nucleon EDM induced by the qCEDM. 

\bibliography{refs}
\end{document}